\documentclass[prb,nofootinbib,twocolumn,superscriptaddress]{revtex4} 

\usepackage{graphicx}
\usepackage{dcolumn}
\usepackage{bm}
\usepackage{threeparttable}
\usepackage{times}
\usepackage{mathptmx}
\usepackage{lscape}
\usepackage{natbib}
\usepackage{amsmath}
\usepackage{amssymb}
\usepackage{braket}
\usepackage{comment}
\usepackage{color}

\def\degree{\kern-.2em\r{}\kern-.3em}

\begin{document}


\title{ Canonical Nonlinearity for Coupled Linear Systems  }

\author{Keigo Kamei}
\affiliation{
Department of Materials Science and Engineering,  Kyoto University, Sakyo, Kyoto 606-8501, Japan\\
}%

\author{Koretaka Yuge}
\affiliation{
Department of Materials Science and Engineering,  Kyoto University, Sakyo, Kyoto 606-8501, Japan\\
}%

\begin{abstract}
{ 
For classical discrete system under constant composition, typically reffered to as substitutional alloys, correspondence between interatomic many-body interactions and structure in thermodynamic equilibrium exhibit profound, complicated nonlinearity (canonical nonlinearity). 
Our recent studies clarify that the nonlinearity can be reasonablly described both by specially-introduced vector field on configuration space and by corresponding diverngence on statistical manifold. 
While these studies shown that the correlation between vector field and local contribution to the divergence can be well characterized by coordination number for a set of selected structural degree of freedoms (SDFs), it is unclear whether the correlations between different set of SDFs purely comes from the difference in covariance matrix of CDOS (determined by coordination number) or   additional information such as the shape of CP should be further required. 
To clarify the problem, we here propose simplified model of the so-called ``Coupled Linear System (CLS)'', which consists of the $m$-mixture of the configurational density of states for linear systems. We demonstrate that the CLS can reasonablly capture the changes in the local nonlinearity w.r.t. the changes in coordination number. Through the dynamic mode decomposition on CLS, we elucidate that there exists two dominant modes to capture the changes in the nonlinearity, where the one uniformly evolves from random to ordered configuration, and the another individually evolves around random, partially ordered and ordered configuration.

}
\end{abstract}


\maketitle

\section{Introduction}
For classical discrete systems under \textit{constant} composition, structural information in thermodynamic equilirbrium along given coordination $p$ can be typically given through the canonical average, namely
\begin{eqnarray}
\label{eq:can}
\Braket{ q_{p}}_{Z} = Z^{-1} \sum_{i} q_{p}^{\left( i \right)} \exp \left( -\beta U^{\left( i \right)} \right),
\end{eqnarray}
where $Z$ denotes partition function, $U$ potential energy, $\beta$ inverse temperature, and summation is taken over all possible microscopic configuration $i$.  
Under the defined corrdination $\left\{ q_{1},\cdots, q_{f} \right\}$, potential energy for any given configuration $k$ can be exactly given by e.g., the linear combination of basis functions for generalized Ising model (GIM):\cite{ce}
\begin{eqnarray}
\label{eq:u}
U^{\left( k \right)} = \sum_{j} \Braket{U|q_{j}} q_{j}^{\left( k \right)},
\end{eqnarray}
where we write inner product, i.e., trace over possible configurations, as $\Braket{\quad|\quad}$. 
These definitions of the equilibrium structure and potential energy naturally induces the canonical average as a map $\phi$:
\begin{eqnarray}
\phi\left( \beta \right): \mathbf{U}  \mapsto \mathbf{Q_{Z}},
\end{eqnarray}
where 
\begin{eqnarray}
\vec{U}&=&\left\{ \Braket{U|q_{1}},\cdots,\Braket{U|q_{f}} \right\} \nonumber \\
\vec{q_{Z}}&=&\left\{ \Braket{q_{1}}_{Z},\cdots,\Braket{q_{f}}_{Z} \right\}.
\end{eqnarray}
Generaly, $\phi$ exhibits complicated nonlinearity w.r.t. $\vec{U}$, resulting in difficulty of exactly predicting equilibrium structure from given potential energy for alloys. Moreover, the nonlinearity straightforwardly relates to the nonlinear behavior for dynamical variables $A$ (e.g., volume and elastic modulus) through $\Braket{A}_{Z} = \sum_{j}\Braket{A|q_{j}}\Braket{q_{j}}_{Z}$. 
Therefore, to understand the nonlinearity between potential energy and equilibrium properties, understandings for $\vec{q_{Z}}-\vec{U}$ nonlinearity should be fundamental.

To address the nonlinearity, we recently introduce ``anharmonicity in the structural degree of freedoms'' (ASDF)\cite{asdf} that is the vector field on configuration space, independent of temperature as well as many-body interactions. We also extend the concept of the nonlinearity based on statistical manifold, successfully decompose ASDF-corresponding nonlinearity as Kullback-Leibler divergence into three contributions in terms of the structural degree of freedoms (SDFs).\cite{ig} 
Based on the above measures, we recently investigate the behavior of the vector field and KL divergence, finding that there exist diverce behavior of ASDF and KL divergence w.r.t. the choice of a set of SDFs, and furthermore, significant positive linear correlation among these measures, whose gradient can be well characterized by the coordination number for pair correlations considered in the given lattice. 
However, for such practical systems, choice of a set of SDFs typically results in deformation of the configurational polyhedra (CP), which can severly affect the NOL. Therefore, from the previous studies, it remains still unclear whether the difference in NOL as vector field, KL divergence and their correlations between different set of SDFs purely comes from the difference in covariance matrix of CDOS (fully determined by coordination number) or further requires additional information such as the shape of CP.

To solve the problem, we here propose simplified model of treating the changes in CDOS with constant shape of CP, by introducing the so-called ``Coupled Linear System (CLS)'', which consists of the $m$-mixture of the CDOS for linear systems. 
We demonstrate that the CLS can qualitatively capture the changes in nonlinearity as the vector field and KL divergence w.r.t. the changes in coordination number as covariance matrix for CDOS. 
By applying the dynamic mode decomposition (DMD) to the successive nonlinearity in CLS, we clarify that the changes in nonlinearity can be characterized by uniformly-evolving mode from random to ordered configuration and individually-evolving mode around random, partially ordered and ordered configurations. The details are shown below.

\section{Concepts and Discussions}
\subsection{Nonlinearity measure}
We first briefly explain the basic concept of the special vector field, ASDF, to define local canonical nonlinearity at given configuration $\vec{q}$:
\begin{eqnarray}
\vec{H}\left( \vec{q} \right) = \left\{ \phi\left( \beta \right) \circ \left( -\beta \Gamma \right)^{-1} \right\}\cdot \vec{q} -\vec{q},
\end{eqnarray}
where $\Gamma$ denotes covariance matrix for CDOS before applying many-body interactions to the system. 
The ASDF has several features of (i) it is independent of temperature and many-body interactions and (ii) when $\phi$ is locally (or globally) linear around $\vec{q}$, $\vec{H}$ takes zero vector. Therefore, ASDF is a natural measure for the local nonlinearity in canonical average of Eq.~\eqref{eq:can}.
Note that ASDF always takes zero at any configuration for LS, whose CDOS takes Gaussian with covariance matrix of $\Gamma$, since $\phi = -\beta \Gamma$ for LS. Note that when we consider discretized Gaussian CDOS, the local linearity inside the configurational polyhedra practically holds (see example at Appendix A), where the discretization is employed so that support of the practical (discrete) CDOS is included in that of discretized Gaussian. 

Then we explain the basic concept of KL divergence $D_{\textrm{KL}}$ as further non-local nonlinearity at configuraion $\vec{q}_{A}$, by extending the concept of the ASDF, which is defined by
\begin{eqnarray}
D_{\textrm{NOL}}^{A} = D_{\textrm{KL}} \left(c_{A}: c_{\textrm{G}A}\right),
\end{eqnarray}
where $c_{A}$ corresponds to canonical distribution under many-body interaction $\vec{V}_{A}=\left(-\beta\Gamma\right)^{-1}\cdot \vec{q}_{A}$ with practical CDOS, and $c_{\textrm{G}A}$ denotes canonical distribution under the same $\vec{V}_{A}$ with CDOS of multidimensional Gaussian with its covarinance matrix takes same as the practical system, i.e., $\Gamma$. 
The nonlinearity as KL divergence can be further decomposed based on the generalized Pythagorean theorem for CDOS with zero covariance, namely, 
\begin{eqnarray}
D_{\textrm{NOL}}^{A} = D_{\textrm{dG}}^{A} + D_{\textrm{NS}}^{A}.
\end{eqnarray}
The former mainly comes from local nonlinearity information exhibiting strong positive correlation with the vector field for practical systems, and the latter comes from further non-local nonlinearity.

\subsection{Coupled Linear System (CLS)}
We have shown that any deviation in CDOS from a set of Gaussin results in nonlinear systems, therefore indicating that proper coupling of the LSs can mimic the nonlinear behavior of practical systems. We first consider the $\alpha$-mixture of probability distributions, which is a natural extention of the concept of statistical manifold ($\alpha=\pm 1$), e.g., for $\alpha$-mixture of discrete distributions $p=\left\{ p_{1},\cdots , p_{k} \right\}$ and $q=\left\{ q_{1},\cdots, q_{k} \right\}$ is given by
\begin{eqnarray}
h_{\alpha}\left( r_{i} \right) &=& \psi\left( c \right) \left\{ \left( 1-c \right) \cdot h_{\alpha}\left( p_{i} \right) + c\cdot h_{\alpha}\left( q_{i} \right) \right\} \nonumber \\
h_{\alpha}\left( s_{i} \right) &=& \frac{2}{1-\alpha} \left\{ \left( s_{i} \right)^{\left( 1-\alpha \right)/2} - 1 \right\},
\end{eqnarray}
where $c$ denotes coupling parameter $\left( 0\le c \le 1 \right)$ and $\psi$ the nomalization factor for probability distribution, and extention to multiple mixture can be straightforward. 
The $\alpha$-mixture is induced by a family of affine connections for Riemannian manifold with statistical invariance w.r.t. random variables for discrete probability distributions: Particularly, $\alpha = -1 \left( +1 \right)$ respectively corresponds to $m$- ($e$-)geodesic on statistical manifold. 
Changes in the value of $\alpha$ can construct variety of CLSs, and we here focus on the case of $\alpha=-1$ (i.e., linear mixture), which is a simple but one of the most fundamental coupling of constituent LSs as follows. 

Among the various $\alpha$-mixtures, only $\alpha=\pm 1$ results in dually flat Riemannian manifold that could be appropriate to treat a set of discrete probability distributions. When we take $\alpha=1$, corresponding to linear mixture of logarism for probability distribution, we can easily find that CLS with $c=0.5$ corresponds to LS, i.e., not only the boundary of $c$, but also intermediate $c$ results in the LS, which is not a desired character since we would like to address the changes in nonlinearity with changes in coordination number as the $c$-dependece of the nonlinearity. 
With these considerations, to investigate how the NOL behaves for coupling of LSs, we here construct CLS with $-1$-mixture of the LS with the same origing and the different variance in CDOS (covariance taking zero). Therefore, its CDOS can be simply given by
\begin{eqnarray}
g_{c}\left(\vec{q}\right) = c\cdot n_{0}\left(\vec{q}\right) + \left(1-c\right)\cdot n_{1}\left(\vec{q}\right),
\end{eqnarray}
where $n_{0}$ and $n_{1}$ respectively corresponds to bivariate Gaussian distribution with its mean taking origin, and their variances are $\sigma_{0} = 0.003$ and $\sigma_{1}=0.01$. 
Note that the present CLS corresponds the following simplified model for practical systems: 
Since for pair correlation CDOS on binary alloys, its origin and variance are respectively determined by composition and coordination number, parameter $c$-dependence of the nonlinearity in CLS sees changes in coordination number at constant composition on a certain given lattice, where increase in $c$ corresponds to increase in the coordination number through decrease in the variance for CLS.

\subsection{Nonlinearity for CLS}
\begin{figure}[h]
\begin{center}
\includegraphics[width=1.03\linewidth]{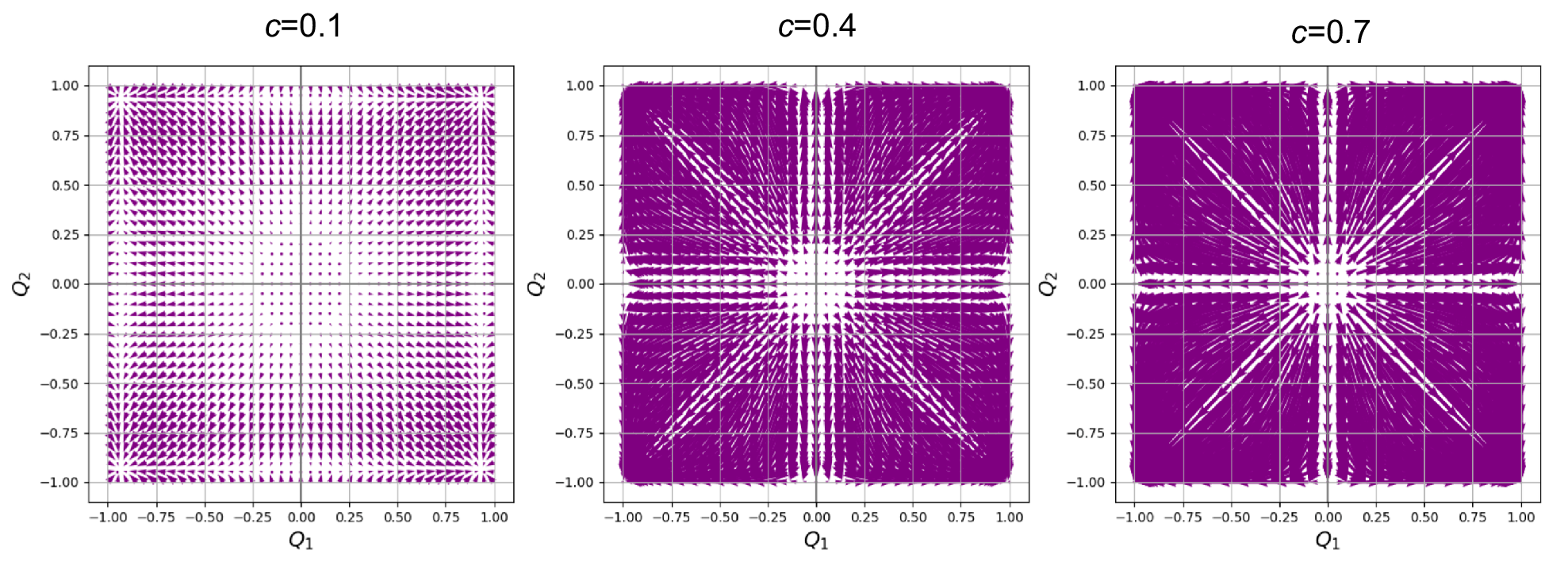}
\caption{ $c$-dependence of the vector field for the CLS. }
\label{fig:ac1}
\end{center}
\end{figure}

We first show in Fig.~\ref{fig:ac1} dependence of ASDF vector field on coupling parameter $c$ for the CLS
We can clearly see that (i) around the origin, the vectors take almost zero for all $c$, due mainly to the locally linear character of $\phi$ around the random configuration, and (ii) the vertices of CP acts as adsorption points, mainly due to that there exist multiple candidates for many-body interaction to provide ground-state configuration, which can be fully characterized by corresponding tangent line for individual CP vertices. These features of the vector field holds for various combination of pair correlations for practical binary systems on representative lattices including fcc and bcc. 

We then address non-local nonlinearity as KL divergence of CLS, where $c$-dependence of individual contribution of $D_{\textrm{dG}}$ and $D_{\textrm{NS}}$ are shown in Fig.~\ref{fig:dkl}. 
\begin{figure}
\begin{center}
\includegraphics[width=1.03\linewidth]{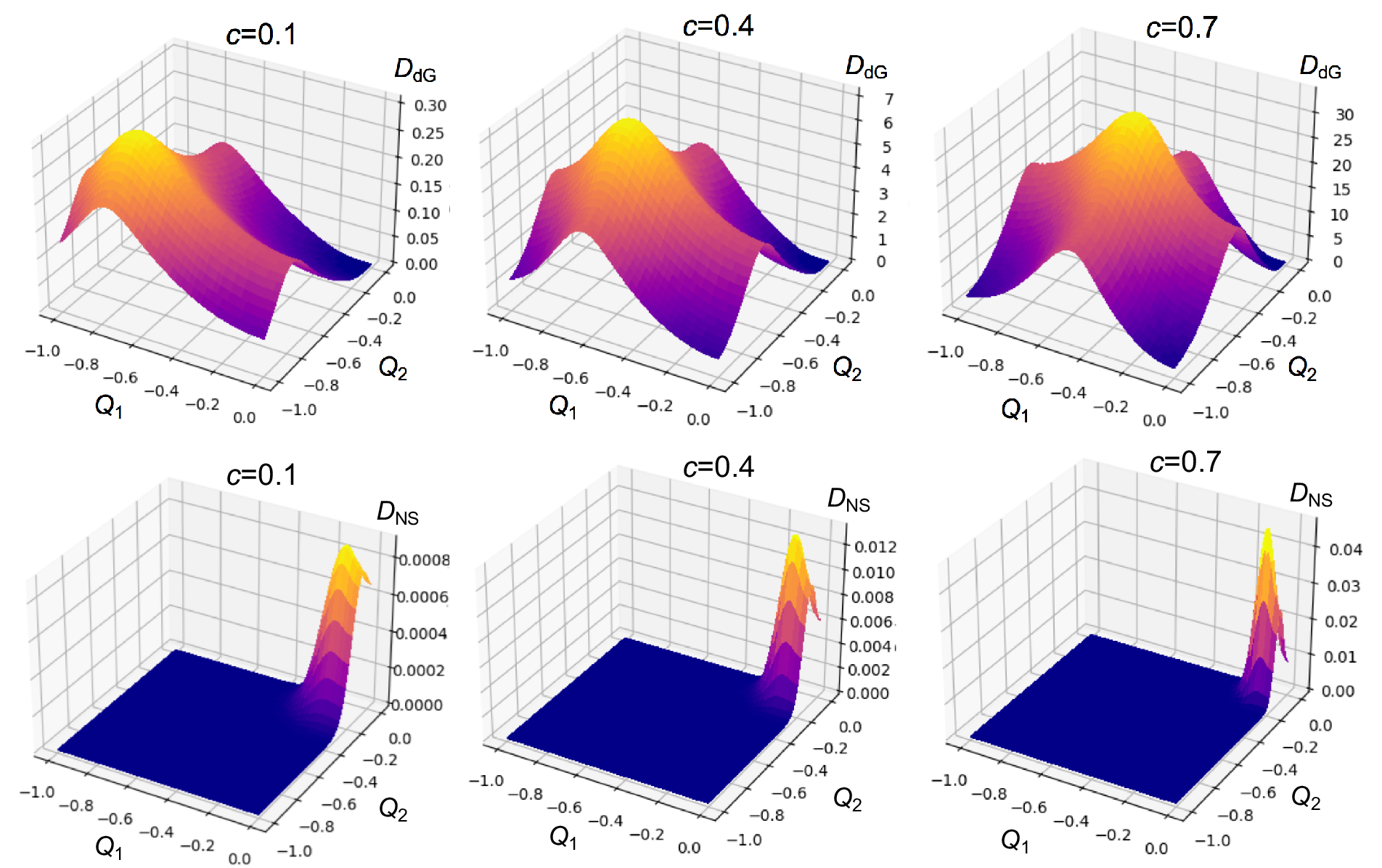}
\caption{ $c$-dependence of non-local nonlinearity as KL divergence for the CLS. }
\label{fig:dkl}
\end{center}
\end{figure}
\begin{figure}[h]
\begin{center}
\includegraphics[width=1.00\linewidth]{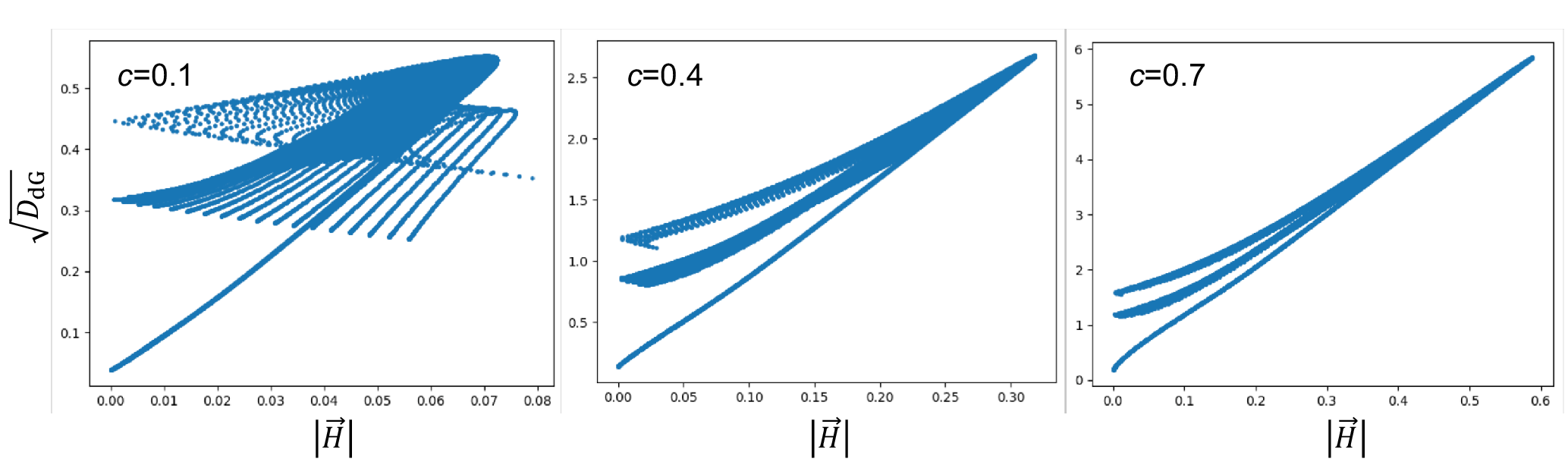}
\caption{$c$-dependence of correlation between $\left|\vec{H}\right|$ and $\sqrt{D_{\textrm{dG}}}$. }
\label{fig:dkl}
\end{center}
\end{figure}
We see the common features of (i) $D_{\textrm{dG}}$ exhibit minimum around the origin and take maxima around partially ordered configuration, and (ii) $D_{\textrm{NS}}$ exhibit maxima around near (but not) origin and take minima around ordered configurations. 
We also show in Fig.~\ref{fig:dkl} the correlation between the vector field and square root of $D_{\textrm{dG}}$. We can see that they exhibit explicit positive correlation, where the gradient increases with increase of $c$: This is consistent with the correlation for practical systems, where they exhibit positive correlaation with its gradient increases with increase of coordination number (corresponding to increasing $c$). For smaller $c$, the scatter for the correlation is more enhanced, while for practical systems, such explicit scatter does not appear: This deviation may come from the difference in the characteristic landscape of CDOS especially around ordered or partially ordered configuration, where the CLS for smaller $c$ tend to overestimate the density of states compared with the practical system, which would result in that variance for canonical distribution of CLS can be much more affected by CP geometry, which has not typically been hold for the practical systems.  

With these discussions, the common features in local (as vector field) and non-local (as KL divergence) nonlinearity between CLS and practical systems would support the validity for the present CLS to mimic the dependecne of nonlinearity on changes in coordination number on real lattice, through the changes in variance of CDOS for the CLS. 
However, there exists distinct difference in the nonlienearity between CLS and practical systems: For CLS, configuration providing maxima for $D_{\textrm{NS}}$ exhibit weak dependence on $c$, while for practical systems, configuration at maxima for $D_{\textrm{NS}}$ typically, significantly depends on a set of SDFs. This fact certainly indicates that strong dependence of non-local nonlinearity landscape w.r.t. a set of SDF cannot be simply characterized by the changes in variance in CDOS, and thus further complicated information for lattice other than coordination number should be required, e.g., the shape of CP.

\subsection{Dominant Contribution to Nonlinearity in CLS}
From above discussions, we see that the CLS can reasonablly capture the characteristic local nonlinearity change w.r.t. the changes in coordination number for practical systems through $c$-dependence of the vector field as well as $D_{\textrm{dG}}$. 
In order to extract which ``modes'' dominates such changes in nonlinearity, we here apply dynamic mode decomposition (DMD) to the $c$-dependence of the vector field, where the DMD has been originally proposed to address spatio temporal evolution of dynamical systems including fuild dynamics. 

\begin{figure}[h]
\begin{center}
\includegraphics[width=0.99\linewidth]{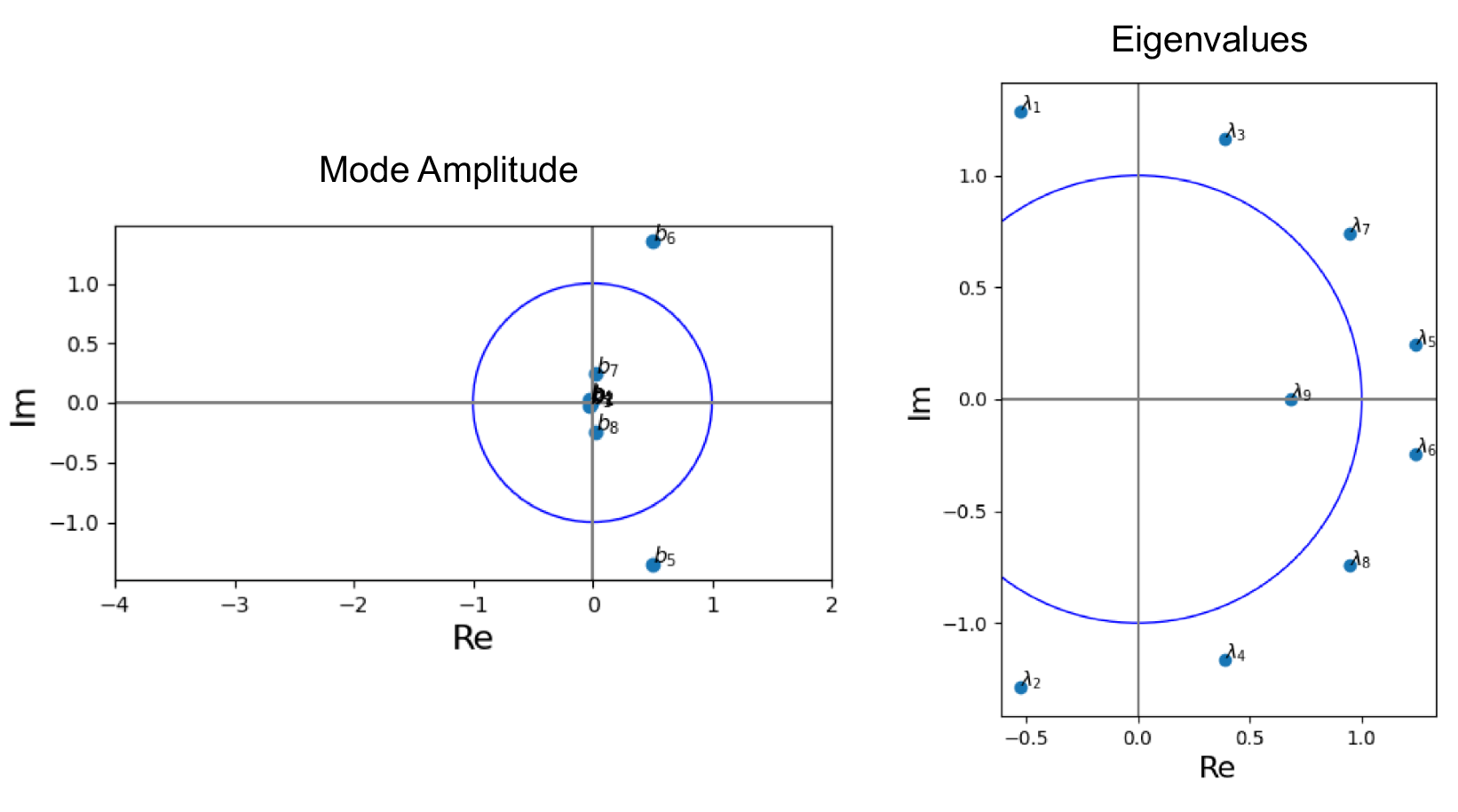}
\caption{DMD mode amplitudes and their eigenvalues on complex plane. }
\label{fig:dmd}
\end{center}
\end{figure}
\begin{figure}[h]
\begin{center}
\includegraphics[width=0.95\linewidth]{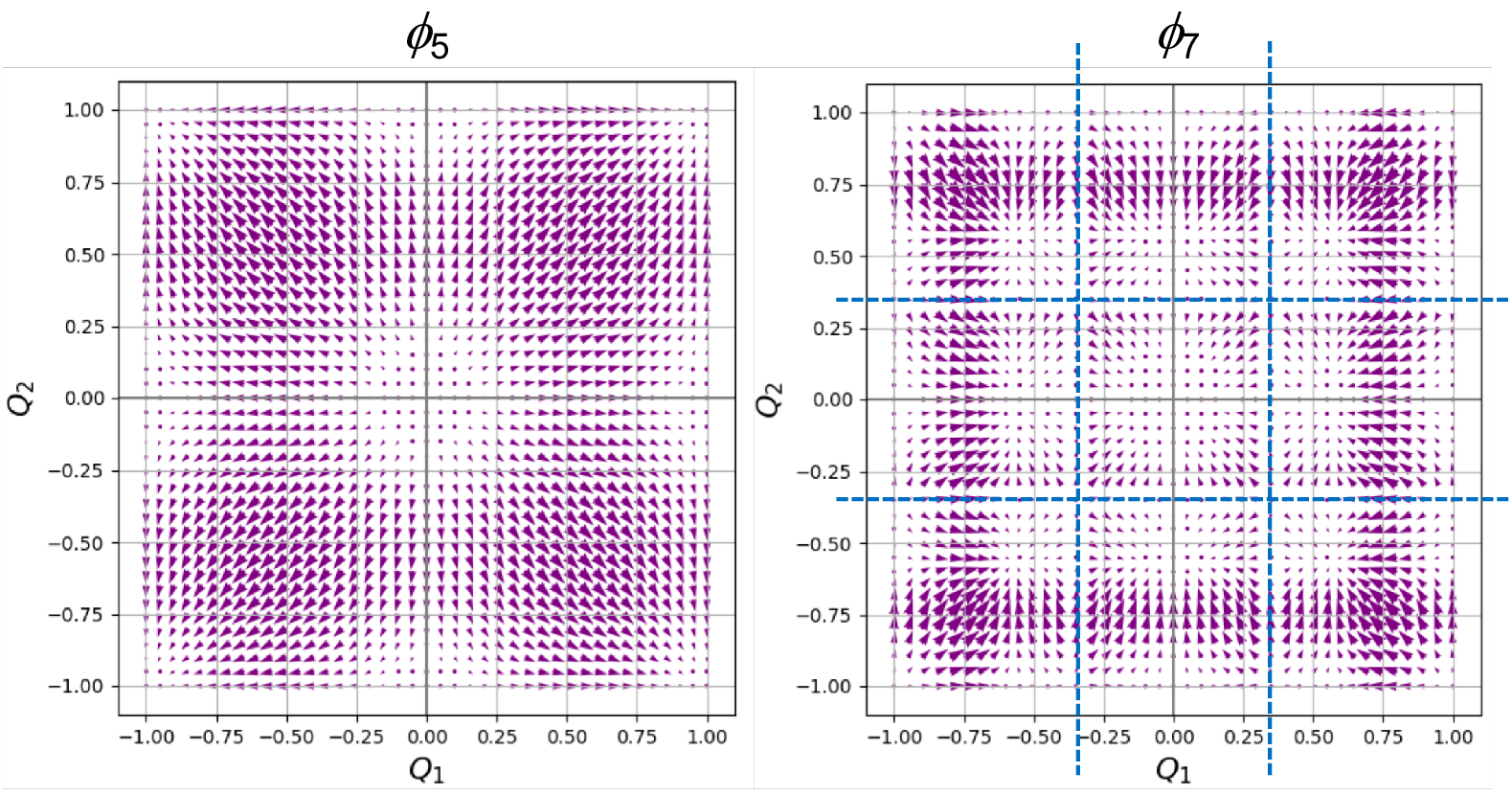}
\caption{Dominant evolving DMD modes. $\phi_{7}$ mode exhibit individual changes in $\vec{H}$ for 9 areas divided by  the 4 dashed lines.}
\label{fig:modes}
\end{center}
\end{figure}

We first briefly explain the DMD. When we define $n$-dimensional column vector $x_{c}$ as the (rearranged) data for vector field $\vec{H}$ over all configurations at parameter $c$ (here, coupling parameter or disntance), we can individually construct parameter series matrix $\mathbf{X}=\left( x_{1},\cdots, x_{m} \right)$ and $\mathbf{X'}=\left( x_{2},\cdots,x_{m+1} \right)$, where $\left( m+1 \right)$ denotes number of snapshots. The DMD consider the best, finite dimensional linear operator $\mathbf{A}$ that satisfies
\begin{eqnarray}
\label{eq:min}
\mathbf{A} = \min \|\mathbf{AX-X'}\|_{\textrm{F}},
\end{eqnarray}
where $\|\cdot\|_{\textrm{F}}$ denotes Frobenius norm. Such $\mathbf{A}$ can be found through
\begin{eqnarray}
\mathbf{A} = \mathbf{X'}\mathbf{X^{\dagger}} = \mathbf{X'V}\Sigma^{-1}\mathbf{U^{*}},
\end{eqnarray}
where $\dagger$ and $*$ respectively denotes Moore-Penrose type pseudo-inverse and adjoint, and $\mathbf{U}$, $\mathbf{V}$ and $\Sigma$ can be obtained through the singular value decomposition of $\mathbf{X}$, i.e., $\mathbf{X} = \mathbf{U}\Sigma\mathbf{V^{*}}$.
Therefore, $\mathbf{A}$ corresponds to the best $n\times n$ matrix to approxmate nonlinear evolution of the vector field w.r.t. changes in given parameter, in terms of Eq.~\eqref{eq:min}. The great advantage to employing the DMD is that we can decompose complicated changes in vector field into individual contribution from DMD mode $\Phi$ (i.e., a set of column eigenvectors for $\mathbf{A}$) and DMD amplitudes $\Lambda$ (i.e., a set of eigenvalues of $\mathbf{A}$). 
Figure~\ref{fig:dmd} shows the resultant DMD mode amplitudes and their eigenvalues. Consider that the modes are comples conjugate, from the figure, we can see that there are two dominant evolving modes indexed by $\sharp 5$ (or $\sharp 6$) and $\sharp 7$ (or $\sharp 9$). 
The corresponding dominant DMD modes are shown in Fig.~\ref{fig:modes}. We can see that $\phi_{5}$ exhibit uniform evolution of the nonlienarity from random (at origin) to ground-state ordered (at vertices) configuration, while $\phi_{7}$ exhibit individual evolution for 9 areas divided by the 4 dashed lines, which can correspond to around random, partially ordered and ordered configurations. We confirm that the uniform and individual (for 9 areas) evoltution can be seen for other CLSs (with other set of variance in CDOS of constituent LSs). These fact strongly indicate that when we address the changes in nonlinearity w.r.t. changes in coordination number, the characteristic nonlinearity individually evolves around random partially ordered and ground-state ordered configurations.

\section{Conclusions}
For classical discrete system under constant composition, we consider the canonical nonlinearity (i.e., nonlinear correspondence between microscopic configuration and many-body interaction under thermodynamic equilibrium) by coupled linear system (CLS) in order to clarify the effect of the changes in variance of CDOS coming from the changes coordination number on the nonlienarity. 
We confirm that the CLS can capture the changes in the local nonlinearity for vector field as well as KL divergence. Applhing the dynamic mode decomposition to CLS nonlinearity, there exists two dominant modes, i.e.,the one uniformly evolves from random to ordered configuration, and the another individually evolves around random, partially ordered and ordered configuration.

\section{Acknowledgement}
This work was supported by Grant-in-Aids for Scientific Research on Innovative Areas on High Entropy Alloys through the grant number JP18H05453 and  from the MEXT of Japan, and Research Grant from Hitachi Metals$\cdot$Materials Science Foundation.

\section*{Appendix A: Linearity for Discretized Gaussian CDOS}
\begin{figure}[h]
\begin{center}
\includegraphics[width=0.6\linewidth]{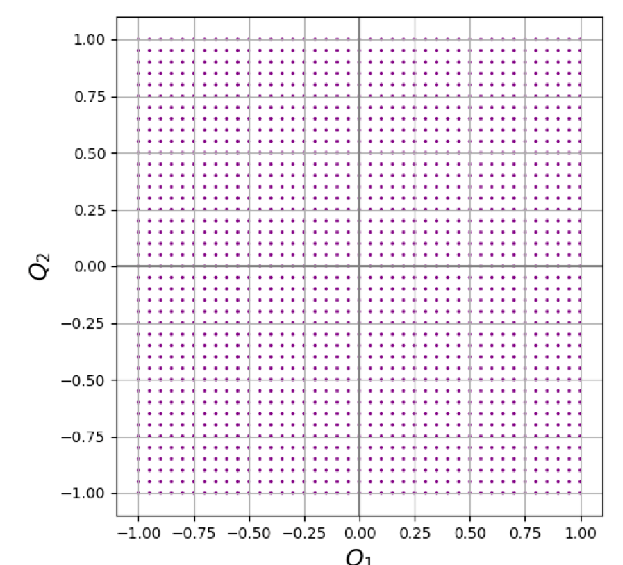}
\caption{ Simulated vector field for the linear system (i.e., $c=0$ with its CDOS taking multidimensional Gaussian). }
\label{fig:zero-vector}
\end{center}
\end{figure}

We here show that local linearity in terms of the vector field can practically hold on inside the configurational polyhedra for linear systems (i.e., its CDOS takes multidimensinal Gaussian), as shown in Fig.~\ref{fig:zero-vector}: We can see that any configuration inside the CP can practically take zero vector.

\section*{Appendix B: NOL for Multiple-Parameter CLS }
Since the complexity for NOL discussed above is expected to be further enhanced for multiple CLSs, we should introduce proper approach to capture the characteristic changes in NOL w.r.t. changes in mutiple coupling (or other) parameters. 
While the DMD effectively provides intuition for complicated changes in NOL, its application is generally restricted to the \textit{single} parameter. However, for the present case of CLSs, we would like to treat multiple mixing of the LSs, naturally leads to treating NOL changes w.r.t. \textit{multiple} parameters. In these cases, we here propose DMD-based approach to systematically address many-body \text{interactions} between coupling parameters that cannot be explained by composits of constituent lower-body interactions. To address this problem, we extend the concept of a single DMD operator $\mathbf{A}$ and state matrix $\mathbf{X}$ to a special set of operators and states as follows:
First, we define \textit{standard} state matrix as
\begin{eqnarray}
\label{eq:std}
\mathbf{X}^{++\cdots +} &=& \left( x_{111\cdots 1}, \cdots,  x_{mmm\cdots m}\right),
\end{eqnarray}
where subscript of each column vector $x_{ij\cdots k}$ denotes the state of $f$ parameters, and superscript $++\cdots +$ denotes that a set of index for state vector increase by 1, i.e., considering simultaneous evolution of all $f$ parameters. 
The reason why we select the standard state as Eq.~\eqref{eq:std} is to unify the state that the multiple parameter-evolution operators act on, as follows. 
From the standard state, we can systematically construct a series of parameter-evolution states denoted by prime on the superscript, namely,
\begin{eqnarray}
\mathbf{X}^{+'+\cdots +} &=& \left( x_{211\cdots 1}, \cdots,  x_{\left( m+1 \right)m\cdots m}\right) \nonumber \\
\mathbf{X}^{++'\cdots +} &=& \left( x_{121\cdots 1}, \cdots,  x_{m\left( m+1 \right)\cdots m}\right) \nonumber \\
\mathbf{X}^{+'+'\cdots +} &=& \left( x_{221\cdots 1}, \cdots,  x_{\left( m+1 \right)\left( m+1 \right)\cdots m}\right) \nonumber \\
&\vdots& \nonumber \\
\mathbf{X}^{+'+'\cdots +'} &=& \left( x_{222\cdots 2}, \cdots,  x_{\left( m+1 \right)\left( m+1 \right)\cdots \left( m+1 \right)}\right).
\end{eqnarray}
Then we can naturally define a set of parameter-evolution operator acting on the standard state, given by
\begin{eqnarray}
\mathbf{A}^{+0\cdots 0}\mathbf{X}^{++\cdots +}  &=& \mathbf{X}^{+'+\cdots +} \nonumber \\
\mathbf{A}^{0+\cdots 0}\mathbf{X}^{++\cdots +}  &=& \mathbf{X}^{++'\cdots +} \nonumber \\
\mathbf{A}^{++\cdots 0}\mathbf{X}^{++\cdots +}  &=& \mathbf{X}^{+'+'\cdots +} \nonumber \\
&\vdots& \nonumber \\
\mathbf{A}^{++\cdots +}\mathbf{X}^{++\cdots +}  &=& \mathbf{X}^{+'+'\cdots +'},
\end{eqnarray}
where superscript $+$ or $0$ on $\mathbf{A}$ respectively denotes the evolution or fix the corresponding parameter. 
These operators can be directly obtained in a similar fashion to the original DMD, by performing e.g., SVD of corresponding state matrix. 

When we introduce the \textit{many-body} interaction for multiple parameters as the partial information about paramater evolution that cannot be treated by combination of composite action of lower-order parameter-evolution operators, we here naturally define the $k$-body interaction for $f$-parameter CLSs as
\begin{eqnarray}
\Delta \mathbf{A}^{\left( k \right)} &=& \mathbf{A}^{\left( k \right)} - \frac{1}{M}\sum_{i=1}^{k}\left( \mathbf{A}^{\left( k \right)}_{i}\overline{\mathbf{A}}^{\left( k \right)}_{i} + \overline{\mathbf{A}}^{\left( k \right)}_{i} \mathbf{A}^{\left( k \right)}_{i} \right) \nonumber \\
M&=&
\begin{cases}
2&\left( k=2 \right) \\
2k&\left( k\ge 3 \right),
\end{cases}
\end{eqnarray}
where
\begin{eqnarray}
\label{eq:mi}
\mathbf{A}^{\left( k \right)} &=& \mathbf{A}^{\overbrace{+\cdots +}^{k} \overbrace{0\cdots 0}^{f-k}} \nonumber \\
\mathbf{A}^{\left( k \right)}_{i} &=& \mathbf{A}^{\overbrace{+\cdots +0_{i}+ \cdots +}^{k} \overbrace{0\cdots 0}^{f-k}} \nonumber \\
\overline{\mathbf{A}}^{\left( k \right)}_{i} &=& \mathbf{A}^{\overbrace{0\cdots 0+_{i}0 \cdots 0}^{k} \overbrace{0\cdots 0}^{f-k}}. 
\end{eqnarray}
Here, $0_{i}$ and $+_{i}$ respectively means that $i$-th superscript takes $0$ or $+$, and the above definition naturally includes non-commutative character of parameter-evolution operators. 
Note that without lack of generality, we choose the first $k$ parameters to be considered just for brief description.
Eq.~\eqref{eq:mi} provides systematic decomposition of the many-body parameter interactions to all possible lower-order operators.
For instance, for 2-body and 3-body inteactions under 3-parameter CLS are respectively given by
\begin{widetext}
\begin{eqnarray}
\Delta \mathbf{A}^{\left( 2 \right)} &=& \mathbf{A}^{++0} - \frac{1}{2}\left( \mathbf{A}^{+00}\mathbf{A}^{0+0} + \mathbf{A}^{0+0}\mathbf{A}^{+00} \right) \nonumber \\
\Delta \mathbf{A}^{\left( 3 \right)} &=& \mathbf{A}^{+++} - \frac{1}{6}\left\{ \mathbf{A}^{++0}\mathbf{A}^{00+}  + \mathbf{A}^{00+}\mathbf{A}^{++0} + \mathbf{A}^{+0+}\mathbf{A}^{0+0} + \mathbf{A}^{0+0}\mathbf{A}^{+0+} + \mathbf{A}^{0++}\mathbf{A}^{+00} + \mathbf{A}^{+00}\mathbf{A}^{0++} \right\}. 
\end{eqnarray}
\end{widetext}
The higher-body inteaction can be further decomposed into lower-body information, in this case, rewriting $\Delta A^{\left( 3 \right)}$ as
\begin{widetext}
\begin{eqnarray}
\Delta\mathbf{A}^{\left( 3 \right)} = \mathbf{A}^{+++} &-& \frac{1}{6}\left\{ \mathbf{A}^{+00}\mathbf{A}^{0+0}\mathbf{A}^{00+} + \mathbf{A}^{+00}\mathbf{A}^{00+}\mathbf{A}^{0+0} + \mathbf{A}^{0+0}\mathbf{A}^{+00}\mathbf{A}^{00+} + \mathbf{A}^{0+0}\mathbf{A}^{00+}\mathbf{A}^{+00} + \mathbf{A}^{00+}\mathbf{A}^{+00}\mathbf{A}^{0+0} + \mathbf{A}^{00+}\mathbf{A}^{0+0}\mathbf{A}^{+00} \right\} \nonumber \\
&-& \frac{1}{6}\left\{ \Delta \mathbf{A}^{++0}\mathbf{A}^{00+} + \mathbf{A}^{00+}\Delta\mathbf{A}^{++0} + \Delta \mathbf{A}^{+0+}\mathbf{A}^{0+0} + \mathbf{A}^{0+0}\Delta\mathbf{A}^{+0+} + \Delta\mathbf{A}^{0++}\mathbf{A}^{+00} + \mathbf{A}^{+00}\Delta\mathbf{A}^{0++} \right\}.
\end{eqnarray}
\end{widetext}


\begin{thebibliography}{9}
\bibitem{ce} J.M. Sanchez, F. Ducastelle, and D. Gratias, Physica A \textbf{128}, 334 (1984).
\bibitem{asdf} K. Yuge, J. Phys. Soc. Jpn. \textbf{86}, 104802 (2018).
\bibitem{ig} K. Yuge, J. Phys. Soc. Jpn. \textbf{91}, 014802 (2022).
\bibitem{st} K. Yuge, arXiv:2103.12414 [cond-mat.stat-mech].
\bibitem{dmd} P. J. Schmidt, J. Fluid. Mech \textbf{656}, 5 (2010).
\end{thebibliography}
\end{document}